\documentstyle[12pt]{article}
\newcommand{\draft}{
        \renewcommand{\baselinestretch}{1.0}%
        \small\normalsize%
}

\draft
\begin{document}
\title{\bf Magnetospheric particle acceleration and
X-ray emission of pulsars}
\author{Sevin\c{c} O. Tagieva$\sp1$
\thanks{e-mail:physic@lan.ab.az},
A\c{s}k\i n Ankay$\sp2$
\thanks{e-mail:askin.ankay@boun.edu.tr},
Arzu Mert Ankay$\sp2$
\thanks{arzu.mert@boun.edu.tr}
\\ \\
{$\sp1$Academy of Science, Physics Institute,} \\
{Baku 370143, Azerbaijan Republic} \\
{$\sp2$Bo\u{g}azi\c{c}i University, Physics Department,} \\
{34342-Bebek, \.{I}stanbul, Turkey} \\
}

\date{}
\maketitle
\begin{abstract}
\noindent The available data on isolated X-ray pulsars, their wind
nebulae, and the supernova remnants which are connected to some of
these sources are analyzed. It is shown that electric fields of neutron
stars tear off charged particles from the surface of neutron star
and trigger the acceleration of particles. The charged particles
are accelerated mainly in the field of magneto-dipole radiation
wave. Power and energy spectra of the charged particles depend on
the strength of the magneto-dipole radiation. Therefore, the X-ray
radiation is strongly dependent on the rate of rotational energy
loss and weakly dependent on the electric field intensity. Coulomb
interaction between the charged particles is the main factor for
the energy loss and the X-ray spectra of the charged particles.
\end{abstract}
Key words: Pulsar, PWN, X-ray

\section{Introduction}
The number of isolated X-ray pulsars and the radio pulsars which
have been detected in X-ray with characteristic age ($\tau$) up
to 10$^6$ yr located at distances less than 7 kpc from the Sun
together with 2 pulsars in Magellanic Clouds is 35 (see Table 1).
The so called magnetars (anomalous X-ray pulsars and soft gamma
repeaters) and dim radio quiet neutron stars like the isolated young
X-ray pulsar 1E 1207.4-5209 are not included in this list as our main
aim in this article is to examine the non-thermal synchrotron
(power-law) emission of isolated young pulsars with relatively well
known ages based on the existing observational data. Also, our list
includes only the pulsars which are certainly not accreting sources.
In order to exclude thermal emission of pulsars, we will particularly
focus our attention on the hard X-ray emission of the pulsars in the
2-10 keV band. On the other hand, we have included the old recycled
millisecond pulsars in the list and examined them in detail throughout
the text to make comparisons between them and the isolated young pulsars.

Most of the pulsars displayed in Table 1 have been observed in 0.1-2.4 keV band $^{1,2}$
and in 2-10 keV band $^3$ and all the available data have been collected, revised
and put together as represented in the table. All of these pulsars
have rate of rotational energy loss \.{E}$>$3$\times$10$^{34}$
erg/s except J1718-3718 for which \.{E} is as low as 1.6$\times$10$^{33}$ erg/s.
Nine of these 35 pulsars are connected to F-type supernova
remnants (SNRs) and 9 of them are connected to C-type SNRs.
Therefore, there exist pulsar wind nebulae (PWNe) near these 18
pulsars which are located up to 7 kpc from the Sun. Four or five of the pulsars are
connected to S-type SNRs (J0538+2817 -- G180.0-1.7 connection is uncertain)
and 2 other pulsars are connected to SNRs
for which the morphological type is not known. Ten pulsars with
detected X-ray emission are not connected to SNRs and 6 of them
have characteristic ages 2$\times$10$^5$$<$$\tau$$<$10$^6$ yr.
Among the 35 pulsars, 2 of them are X-ray pulsars from which radio
emission has not been observed and they are connected to SNRs
(pulsar J1846-0258 to SNR G29.7-0.3 and pulsar J1811-1925 to SNR
G11.2-0.3). Radio pulsar J1646-4346 (d=4.51 kpc) which is
connected to C-type SNR G341.2+0.9 but not observed in X-ray is
also included in the list. Other than these sources, we have also
included 14 radio pulsars with \.{E}$>$4$\times$10$^{35}$ erg/s,
\.{P}$>$10$^{-15}$ s/s and d$\le$7 kpc in Table 1. Also, 6 isolated
millisecond pulsars from which X-ray radiation have been observed
are displayed in Table 1.

Our main aims in this work are to analyze the conditions which are
necessary for the acceleration of relativistic particles, the
X-ray radiation produced in the magnetospheres of pulsars, and the
presence or absence of PWNe around these objects.

\section{Dependence of the ejection of relativistic particles on different
parameters of pulsars} As known, very young pulsars have the
capability to eject relativistic particles. Power and spectra of
the particles must be related to different parameters of pulsars,
particulary to the induced electric field intensity for which it
is practically not possible to find an exact expression as there
are significant uncertainties in the magnetic field structures of
neutron stars in plasma. Therefore, we use the simplified relation
\begin{equation}
E_{el} = \frac{B_p R}{P}
\end{equation}
$^4$ which must be correct in the case when the braking of
pulsar's rotation is connected to magneto-dipole radiation only. Here,
R is the radius, B$_p$ is the magnetic field
intensity at the magnetic pole, and P is the rotation period of the
neutron star. The number and energy spectra of the relativistic
particles also depend on the conditions in the atmospheres of
neutron stars (density and temperature). Independent of the
mechanism, the rate of rotational energy loss (\.{E}) depends only
on the moment of inertia (I), the rotation period P and the time
derivative of P (\.{P})
\begin{equation}
\dot{E}=\frac{4\pi ^2 I \dot{P}}{P^3}.
\end{equation}

If we assume that practically the rate of rotational energy loss
is related only to the magneto-dipole radiation $^{4,5}$,
then the effective magnetic field
B=B$_p$$sin\alpha$ (where $\alpha$ is the angle between the
surface dipole magnetic field axis and the rotation axis) and
\begin{equation}
\dot{E} = \frac{2}{3} \frac{\mu^2_{\perp} \omega^4}{c^3} =
\frac{8\pi^4}{3c^3} \frac{B^2R^6}{P^4}
\end{equation}
where $\mu_{\perp}$=BR$^3$/2 is the component of the
magnetic moment perpendicular to the rotation axis, $\omega$=2$\pi$/P
the angular velocity and c the
speed of light. From expressions (1) and (3) we get
\begin{equation}
\dot{E} = \frac{8\pi^4}{3c^3} \frac{R^4}{P^2} E_{el}^2
sin^2\alpha.
\end{equation}
As seen from this equation, \.{E} is directly proportional to E$_{el}$$^2$
and inversely proportional to P$^2$.

From expressions (1)-(3), we can get
\begin{equation}
E_{el} \sim I^{1/2} R^{-2} \tau ^{-1/2}
\end{equation}
(where the characteristic age $\tau$=P/2\.{P}) from which it
follows that the lines of E$_{el}$=constant roughly pass parallel
to the lines of $\tau$=constant on the P-\.{P} diagram for fixed
values of I and R. Dependence of $\tau$, E$_{el}$, B, and \.{E} on
the values of I, R, and the braking index (n) and also the
evolutions of pulsars with very different parameters are examined
in detail by Guseinov et al. $^{6,7}$.

The existing theories do not give us the possibility to estimate
the value of the voltage better than the order of magnitude. But
it is exactly known that charged particles can be pulled out from
the hot atmosphere of pulsar because of the large value of
E$_{el}$ and they can be additionally accelerated. Further acceleration
can take place in the field of the magneto-dipole wave $^4$.
Which of these mechanisms is mainly responsible for the
acceleration must be determined from the observational data. It
may be the case that the voltage creates only basis for further
acceleration of particles similar to the case of the Maxwell tail
of hot particles (or particles which have already been accelerated
with some other mechanism) in the shock fronts of SNRs. In the
regular acceleration models $^{8-10}$, a
further acceleration of particles with high energy takes place
under the crossing through the shock fronts of SNRs $^{11}$.
Therefore, it is necessary to analyze the observational data.

\section{How are the observational data related to the theoretical idea?}
From the observations of the radio and X-ray radiation of neutron
stars with power-law spectra (excluding the cooling radiation of
neutron stars), there is direct evidence for continuing particle
acceleration. Such X-ray radiation have been observed, for
example, from pulsars J1856+0113 and J1801-2451 which have PWNe
and have values of \.{E}$\sim$(4-5)$\times$10$^{35}$ erg/s and
2.5$\times$10$^{36}$ erg/s, respectively. On the other hand,
old millisecond pulsars J1939+2134 and J1824-2452 with
\.{E}$\sim$(1-2)$\times$10$^{36}$ erg/s also have similar X-ray
radiation as the previous pulsars $^2$,
but with smaller luminosity in the 2-10 keV band. As seen in Table 1,
these pulsars, beginning from J1856+0113, have values of X-ray
radiation 10$^{33}$, 1.5$\times$10$^{33}$, 5$\times$10$^{32}$, and
3.6$\times$10$^{33}$ erg/s, respectively, in the 2-10 keV band
$^{2,3}$. We have estimated the ratio
($\frac{\dot{P}}{P}$)$^{1/2}$$\sim$$\tau$$^{-1/2}$ from the
dependence of E$_{el}$ (see expression (5)) for pulsars
J1856+0113, J1801-2451, J1939+2134, and J1824-2452 and we have
found the values 9$\times$10$^{-7}$, 10$^{-6}$,
7$\times$10$^{-9}$, and 3$\times$10$^{-8}$ s$^{-1/2}$,
respectively. If in all these four cases the values of radius and
moment of inertia of the pulsars are approximately the same, then
the values of the electric field intensity E$_{el}$
(see expression (5)) may also have
approximately the same ratios. In other words, electric fields of
old millisecond pulsars are more than 2 orders of magnitude smaller
compared to young pulsars with similar L$_{2-10keV}$. \.{E} values of
these four pulsars are roughly on the same order of magnitude, but
the initial energy of electrons is different because their
E$_{el}$ values are different. How is it possible that the
L$_{2-10keV}$ values of these two types of pulsars are comparable
to each other? Young pulsars have PWNe around them, i.e. an extra
X-ray radiation of the electrons in the PWN comes from the source
in addition to the X-ray radiation of the neutron star. These old
millisecond pulsars have log$\tau<$8.37.

In Figure 1, dependence of the X-ray luminosity in 2-10 keV band
(L$_{2-10keV}$) on the value of \.{E} for the isolated pulsars with
comparably hard spectra which are located up to 6 kpc from the Sun
together with 2 pulsars in the
Magellanic Clouds (see Table 1) is represented. As seen in Figure
1, pulsars J1856+0113 and J1801-2451 have \.{E} values similar to
isolated old millisecond pulsars J1824-2452 and J1939+2134, and their
\.{E} values are more than 300 times greater than their X-ray
luminosity in 2-10 keV band. Therefore, the luminosity of the
power-law X-ray radiation (which includes the radiation of the
relativistic particles in the Coulomb and magnetic fields) mainly
depends on the value of \.{E}. How can this approach be confirmed
also for other pulsars based on the observational data?

Since we are interested in the acceleration of particles, we have
chosen the pulsars for which
L$_{2-10keV}$/L$_{0.1-2.4keV}$$>$1, so that, we avoid the errors connected
to the interstellar absorption and the radiation related with the cooling
of neutron stars. As seen from Table 1, these conditions are not
satisfied for the last 4 pulsars with log $\tau$=9.58-9.86 which are also
displayed in Figure 1. But they are very close to the Sun and very old,
therefore they do not create any difficulties. All the young pulsars
shown in Figure 1 have not only hard spectra but also have PWNe. As seen
from the equations of the best fit
\begin{equation}
L_{2-10keV} = 1.42\times10^8 \dot{E}^{0.63}
\end{equation}
for the pulsars with \.{E}$<$10$^{35}$ erg/s and L$_{2-10keV}$$<$10$^{31}$
erg/s and
\begin{equation}
L_{2-10keV} = 1.86\times10^{-37} \dot{E}^{1.92}
\end{equation}
for the other pulsars with higher L$_{2-10keV}$ and/or higher \.{E},
there exist clear relations between \.{E} and L$_{2-10keV}$ for
which the radiation is related with the accelerated particles.

In Figure 2, the relation between L$_{2-10keV}$ and the
characteristic age of the same pulsars shown in Figure 1 is
represented. As seen from the figure, a single dependence for both
the young and the old pulsar populations does not exist (see also
Possenti et al. $^3$). Pulsars with similar values of
L$_{2-10keV}$ and \.{E} may have very different values of
characteristic age and this leads to different dependence for
each population. As their ages increase, the young pulsars
continue to move practically along the same direction on the
L$_x$-$\tau$ diagram. This is seen from Table 1 where there are isolated
pulsars with characteristic ages 10$^5$-10$^7$ yr and with luminosity
in 2-10 keV band similar to the luminosity of the old millisecond pulsars.
Therefore, the electric field intensity does
not determine the intensity and possibly also the spectra of the
relativistic particles. The X-ray luminosity strongly depends on
the value of \.{E} and weakly depends on $\tau$ and E$_{el}$.
Large values of E$_{el}$ mainly trigger the acceleration of
particles.

As seen from Figure 1 and expressions (6) and (7), as \.{E} decreases
the X-ray luminosity of pulsars also decreases. So, in order to have
the probability to observe X-ray radiation from a pulsar to be
high for a fixed value of E$_{el}$, the rotation period of the pulsar must
be short (see expression (4)). On the other hand, as seen from
Figure 1 and Table 1, the L$_x$ values of the young pulsars which
are connected to SNRs are $\sim$5 orders of magnitude larger than
the L$_x$ values of old ms pulsars and this is also seen from
expressions (6) and (7). This must be the result of small number of
relativistic particles in the magnetospheres of old ms pulsars.

As seen from Figure 2, the difference in the $\tau$ values of very old ms
pulsars ($\tau$$>$10$^9$ yr) and young pulsars is about 5.5 orders of
magnitude (and from expression (5) the difference in the E$_{el}$$^2$
values is also about 5.5 orders of magnitude). The differences in the
\.{E} values is about 3-3.5 orders and in the P values
is on average 2.5 orders of magnitude. Considering that the expression
for E$_{el}$ can only be roughly determined as mentioned above, these
data show that the relation between \.{E}, E$_{el}$ and P in expression
(4) is valid.

In order to tear off particles from the surface of neutron star it
is necessary to have large values of E$_{el}$. The occurrence of
this process must be easier under the existence of hot and
extended atmosphere. Therefore, the values of mass and age of
pulsars must also have important roles in the formation of the
X-ray emission.

If the braking index n is constant along all the evolutionary
tracks with different initial parameters, then the age of the
pulsar is
\begin{equation}
t = \frac{P}{(n-1)\dot{P}} (1-(\frac{P_0}{P})^{n-1}) = \tau \times
(1-(\frac{P_0}{P})^{n-1}).
\end{equation}
For the young pulsars P$_0$ (the initial spin period of pulsar) can
assumed to be much less than P, so
that, $\tau$ can approximately be equal to the age. For recycled ms
pulsars, the P value can be comparable to the P$_0$ value (the period
which the radio pulsar is born with after completing the X-ray binary phase,
see Bisnovatyi-Kogan and Komberg $^{12,13}$ on recycled ms pulsars),
so that $\tau$ can be several times larger than the age. For our purposes
such a difference is not significant since it is enough to know the order
of magnitude of the age which is much longer than the cooling time of
recycled pulsars.

As the last 4 pulsars displayed in Table 1 with \.{E} values close to
3$\times$10$^{33}$ erg/s can emit X-ray, all of the isolated
old ms pulsars with \.{E}$>$3$\times$10$^{33}$ erg/s must also have X-ray
radiation. On the other hand, all the 5 old ms pulsars with P$\le$6 ms
which are located up to $\sim$0.5 kpc $^{14,15}$ have been detected
in X-ray $^{2,3}$. Since the number density of old ms
pulsars is very large
compared to the number density of low mass X-ray binaries, the ages
of these pulsars must not be much less than $\tau$ in general. As we see,
although the recycled millisecond pulsars are very old, they still emit
X-ray because they have large \.{E} values.

The X-ray luminosity of pulsars together with their PWNe are
represented in Table 1. Often, such X-ray luminosity values
include both the X-ray radiation coming from the pulsar and the
PWN together, because it is difficult to separate the emission of
the pulsar from the emission of the PWN. Therefore, the radiation
of isolated ms pulsars must be smaller compared to isolated young
pulsars for equal values of \.{E}. This must be true also because
of young pulsars having large values of electric field intensity
and possibly having smaller masses in some cases. Therefore, it is
strange that ms pulsar J1824-2452 has larger luminosity compared
to young pulsar J1801-2451 (see Figure 1). On the other hand,
their radiation is very hard. Actually, the uncertainties in the
luminosity values of these ms pulsars may be large.

In Figure 3, P-\.{P} diagram of all the pulsars in Table 1, which
have connections with SNRs and/or from which X-ray radiation has
been observed, is displayed. As seen from this figure and Table 1,
pulsars Geminga and J0538+2817 which is connected to SNR
G180.0-1.7 $^{16}$ have about one order of magnitude
larger values of \.{E} compared to old ms pulsars, but their
luminosity in 2-10 keV band is similar to the luminosity of these
ms pulsars. The X-ray spectra of the mentioned young pulsars are
steeper compared to the spectra of old ms pulsars as seen from the
comparison of the luminosity in 0.1-2.4 keV and 2-10 keV bands.
The luminosity of J0826+2637 and J0953+0755 in 2-10 keV band are
smaller than the luminosity of old ms pulsars, but their \.{E}
values are also smaller (see Fig.3 and Fig.1). How can this
observational fact be explained?

Young pulsars have values of electric field intensity about 50
times larger than the values of old ms pulsars, they are hotter
and they may also have smaller masses. These facts create
conditions for tearing off charged particles more easily from
neutron stars. On the other hand, the magnetic field values of
young pulsars are more than 3 orders of magnitude larger (the high
magnetic field does not let the particles to escape from the
magnetosphere without considerable radiative energy loss).
These and the possible existence of large number
of charged particles in the magnetospheres and the surroundings of
young pulsars must create the best conditions for magneto-braking
and Coulomb radiation. Among young pulsars, the small values of
X-ray radiation of J0538+2817 and Geminga in 2-10 keV band seem
to contradict this natural discussion.

\section{Pulsar wind nebulae and pulsars in supernova remnants}
Let us now examine the presence or absence of PWN around pulsars
and the X-ray radiation. All the pulsars with $\tau$$<$10$^7$ yr
represented in Table 1 (except pulsar J0537-6910 in the Magellanic
Cloud) are shown on the P-\.{P} diagram in Figure 4 denoting also
the morphological types of the SNRs which some of the pulsars are
connected to. If the type of the SNR is C (composite) or F
(filled-center), then it means that there is PWN created by the
neutron star. In S (shell) type SNRs there is no observed PWN
possibly because it is very faint. Other pulsars in Table 1 which
are not connected to SNRs and which have $\tau$$<$10$^7$ yr are
also displayed in Figure 4. All the pulsars with PWN around them
have L$_x$ (2-10 keV) $>$ 10$^{33}$ erg/s (see Table 1).
The cooling radiation of these pulsars do not have a considerable role.

Pulsars J1803-2137, J1016-5857, J2337+6151, and J0659+1414, which are
connected to S-type SNRs (in the radio band), have been observed to
radiate in X-ray band. Pulsars J2337+6151 and
J0659+1414 have suitable distance values and positions to
observe possible PWNe around them, but \.{E} and L$_x$ (2-10 keV)
values of these pulsars are so small that it is not so likely that
PWNe with observable brightness are present around them.

From the analysis of these data, we see that lifetime of the X-ray
radiation produced by the relativistic particles in the magnetospheres of
neutron stars is considerably longer than lifetime of PWNe (see Table 1). But
there may be one exception, namely pulsar J0538+2817, which is
connected to SNR G180.0-1.7 (S147). This SNR is S-type in radio band
$^{17}$. Romani \& Ng $^{16}$ claim evidence for a faint nebula around
the pulsar, but in a more recent work $^{18}$ no evidence
has been found of a PWN and the pulsar has been observed to radiate only
thermally. Also, the position and ages of the pulsar and the SNR show
that a physical connection between these two sources is dubious and there
are no data directly showing evidence for the connection.

From Figure 4 and Table 1 we see that PWN may exist around pulsars with
\.{E}$>$10$^{35}$ erg/s and $\tau$$<$5$\times$10$^4$ yr. Pulsar
J1646-4346 in C-type SNR G341.2+0.9 satisfies these criteria, but X-ray
radiation has not been observed from it because it is located at a
distance of 6.9 kpc $^{19}$. X-ray radiation has
not been observed also from 9 far away pulsars with d$\ge$3 kpc which are
located close to the Galactic plane in the Galactic central directions
(see Table 1 and Figure 4) and which satisfy the conditions of large
value of \.{E} and small value of $\tau$. As seen from the position of
pulsar J1617-5055 in Figure 4, it must have a connection with SNR and
must have PWN. This follows also from the large values of X-ray
luminosity in both X-ray bands. Lack of observed SNR and PWN
must be related with the large distance and the position in the Galaxy.
Ages of the relatively nearby pulsars which are displayed in Figure 4 may
be smaller than the values of $\tau$. Three of these pulsars are
connected to S-type SNRs and all of them have small X-ray luminosity.

\section{Conclusions}
From the analysis of all the available data on isolated X-ray pulsars, their
wind nebulae and the SNRs which are connected to some of these sources and
also from theoretical considerations, we have found the conclusions below:
\\
1. The electric field intensity (voltage) of neutron stars is not the main
physical quantity for the spectra, energy, and intensity of the
relativistic
particles which produce the X-ray radiation of isolated pulsars and their
wind nebulae. The voltage mainly triggers the acceleration of particles.
In order to tear off charged particles from the surface of neutron star
E$_{el}$ must be large. This process occurs more easily in hot and
extended atmospheres. Therefore, the values of mass and age of pulsars
also have important roles in the formation of the X-ray radiation. \\
2. The acceleration of relativistic particles mainly takes place
in the field of the magneto-dipole radiation wave. Because of
this, the X-ray radiation of pulsars and their wind nebulae
strongly depend on the value of the rate of rotational energy loss
which is reflected by the spectra of the
magneto-dipole radiation. \\
3. The high magnetic field and particularly the number density of
the charged particles create conditions under which the energy
loss of the particles in 0.1-2.4 keV band may surpass the energy
received from the magneto-dipole radiation wave for further
acceleration. This must be responsible for the steeper X-ray
spectra of young pulsars as compared to old ms pulsars. \\
4. PWN exists only around pulsars with \.{E}$>$10$^{35}$ erg/s and
$\tau$$<$5$\times$10$^4$ yr. Also, the X-ray luminosity of pulsars
in 2-10 keV band drops down to $\sim$10$^{33}$ erg/s at about the
same time when the PWN becomes unobservable in both radio and
X-ray bands. The lifetime of S-type SNRs exceeds the lifetimes of
F-type and C-type SNRs. So, it may be possible that C-type SNRs
can show themselves as S-type after some time in their evolution.
PWN must be observed around pulsar J1617-5055. \\ \\
\textit{Acknowledgments} S.O.T is supported by INTAS. A.A. and A.M.A.
thank Bogazici University Foundation (BUVAK).

\clearpage
\begin{figure}[t]
\vspace{3cm} \includegraphics{mass_of_ns_table_GENEL.ps}
\end{figure}

\clearpage
\begin{figure}[t]
\vspace{3cm} \includegraphics{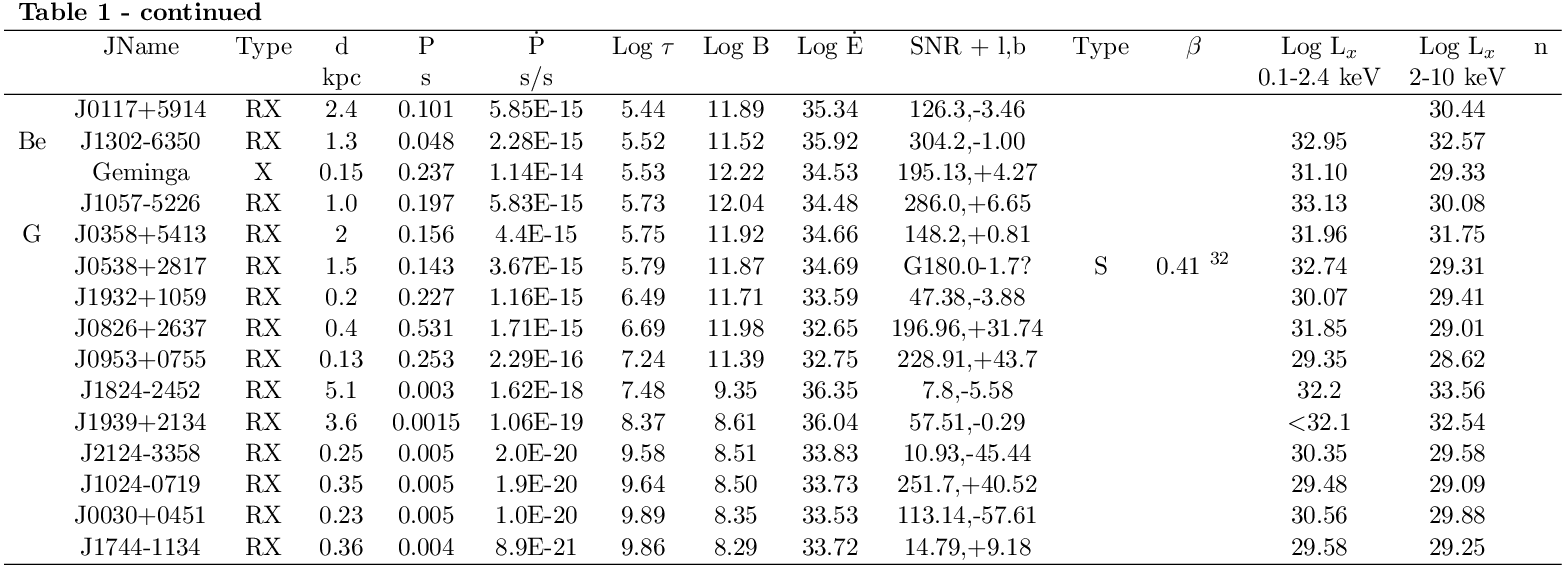}
\end{figure}

\clearpage
{\bf Figure Captions} \\
{\bf Figure 1:} Dependence of the 2-10 keV luminosity on the rate of
rotational energy loss for pulsars located up to 7 kpc from the Sun
including also 2 pulsars in Magellanic Clouds which are denoted by '+'
sign. Old millisecond pulsars are shown by 'star' sign and young
pulsars by 'X' sign. \\
{\bf Figure 2:} Dependence of the 2-10 keV luminosity on the characteristic
age for pulsars located up to 7 kpc from the Sun including also 2
pulsars in Magellanic Clouds. The pulsars are displayed by the same
symbols as in Figure 1. \\
{\bf Figure 3:} Period versus period derivative diagram for the pulsars
which have been detected in X-ray and/or which are connected to SNRs (see
Table 1). The sign '+' represents the pulsars which have physical
connections with SNRs and the 'X' sign denotes the pulsars which are
not connected to SNRs. B8-B14, E29-E41 and T3-T10 represent the constant
lines of the effective surface magnetic field (B=10$^8$-10$^{14}$ Gauss),
the rotational kinetic energy loss (\.{E}=10$^{29}$-10$^{41}$ erg/s) and
the characteristic age ($\tau$=10$^3$-10$^{10}$ yr). \\
{\bf Figure 4:} Period versus period derivative diagram for all 48 pulsars
represented in Table 1 with $\tau$ $<$ 10$^6$ yr and distance $\le$ 7 kpc
including the 2 pulsars in Magellanic Clouds. '+' signs represent the 11
pulsars connected to C-type SNRs, 'X' signs show the 9 pulsars connected
to F-type SNRs, and 'star' signs denote the 4 pulsars connected to S-type
SNRs. 'Light square' signs represent the 14 pulsars which are not connected to
SNRs and not observed in X-ray. The 10 pulsars which are denoted with
'dark squares' have been observed in X-ray but they are not connected to SNRs.
B12-B14, E32-E38 and T3-T6 represent the constant lines of the effective surface
magnetic field (B=10$^{12}$-10$^{14}$ Gauss), the rotational kinetic energy loss
(\.{E}=10$^{32}$-10$^{38}$ erg/s) and the characteristic age
($\tau$=10$^3$-10$^6$ yr). \\

\clearpage
\begin{figure}[t]
\vspace{3cm}
\includegraphics{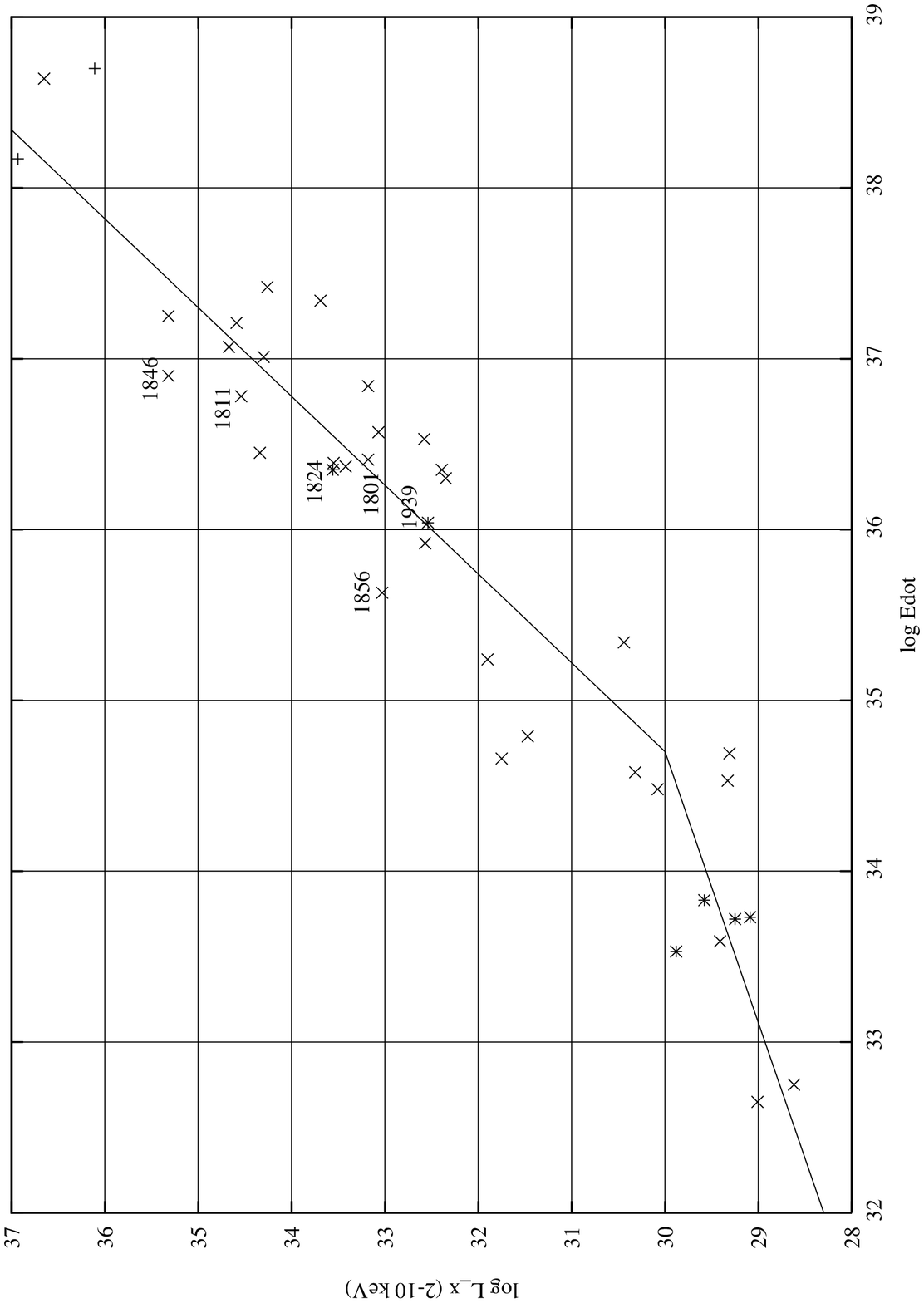}
\end{figure}

\clearpage
\begin{figure}[t]
\vspace{3cm}
\includegraphics{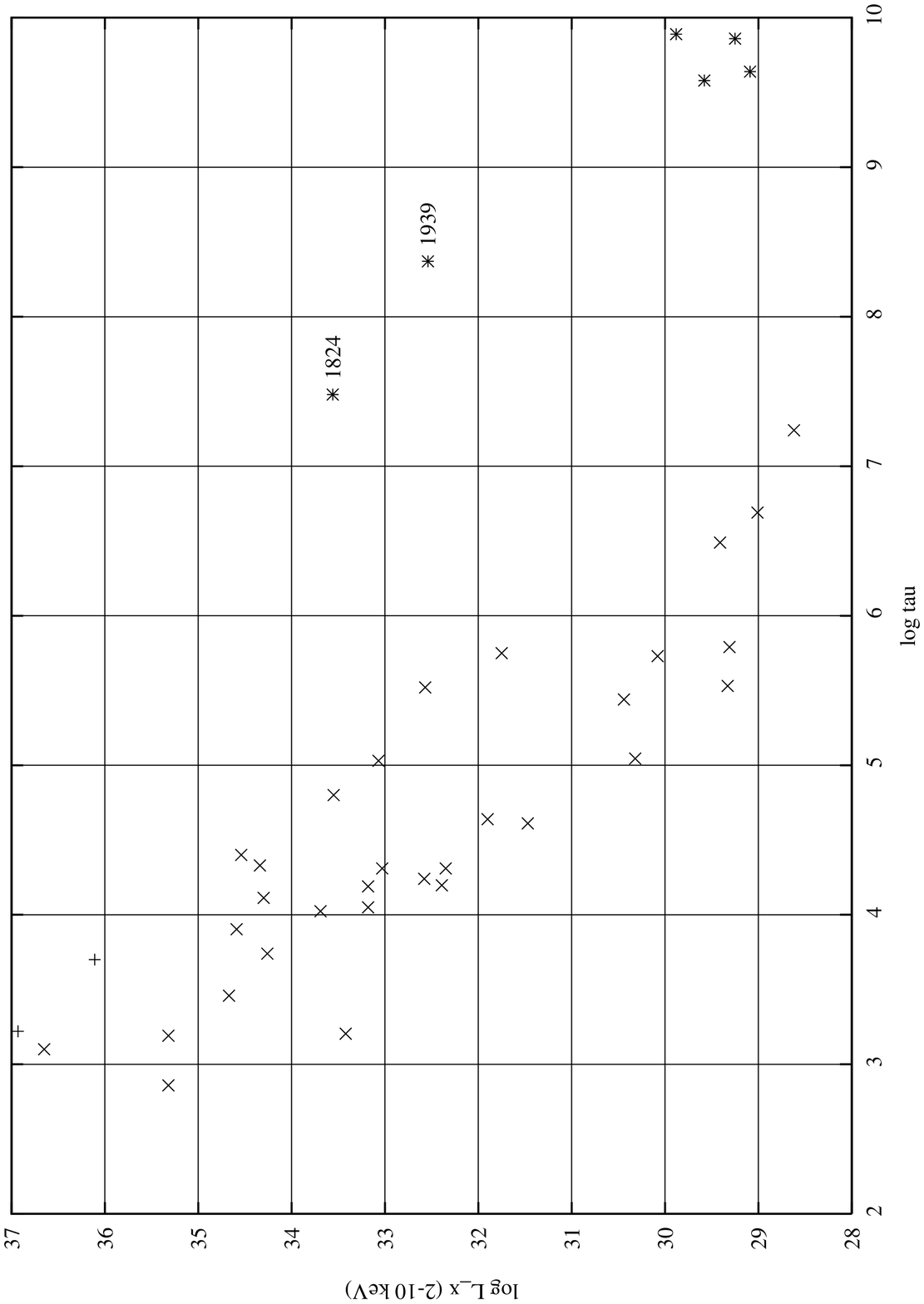}
\end{figure}

\clearpage
\begin{figure}[t]
\vspace{3cm}
\includegraphics{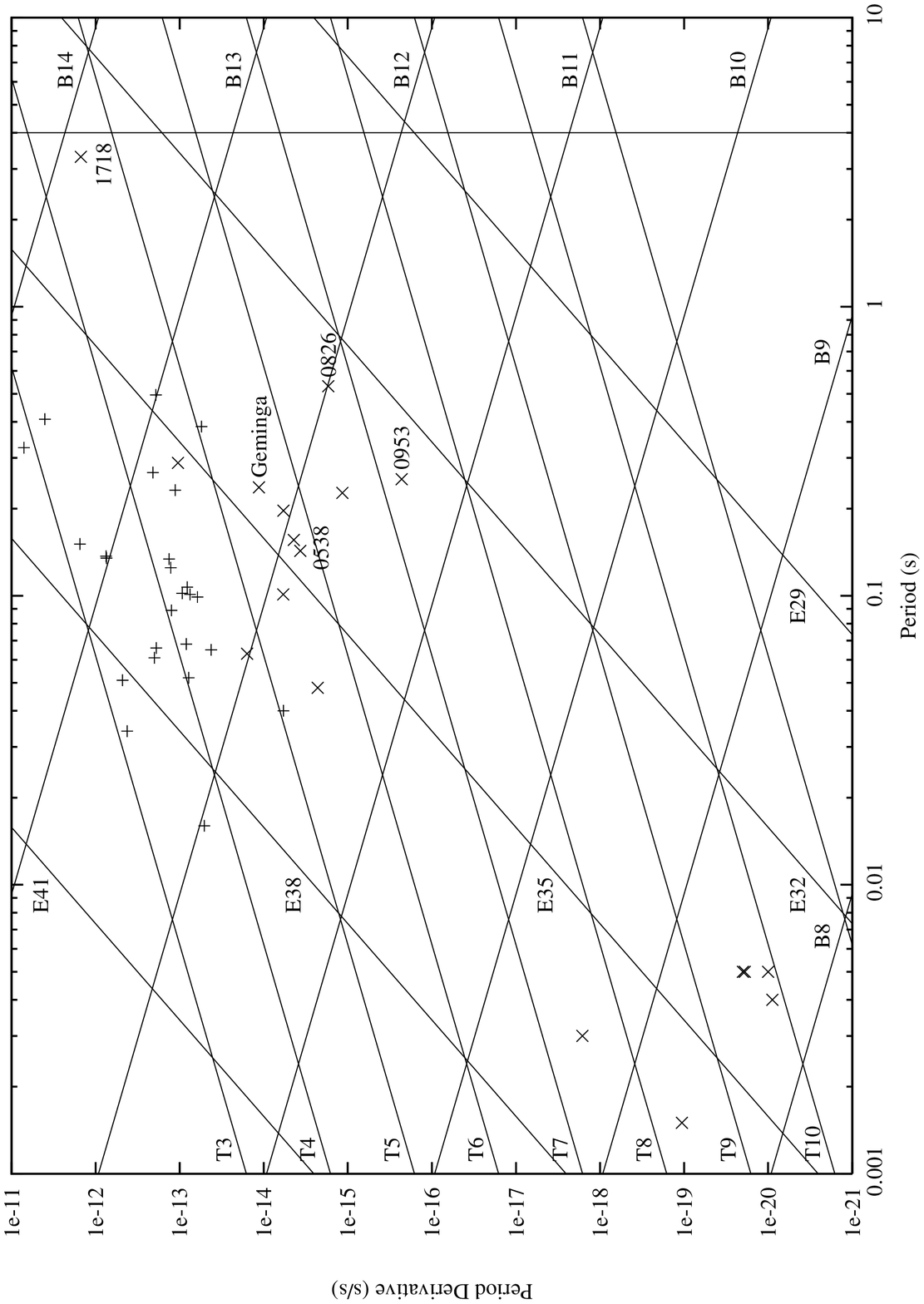}
\end{figure}

\clearpage
\begin{figure}[t]
\vspace{3cm}
\includegraphics{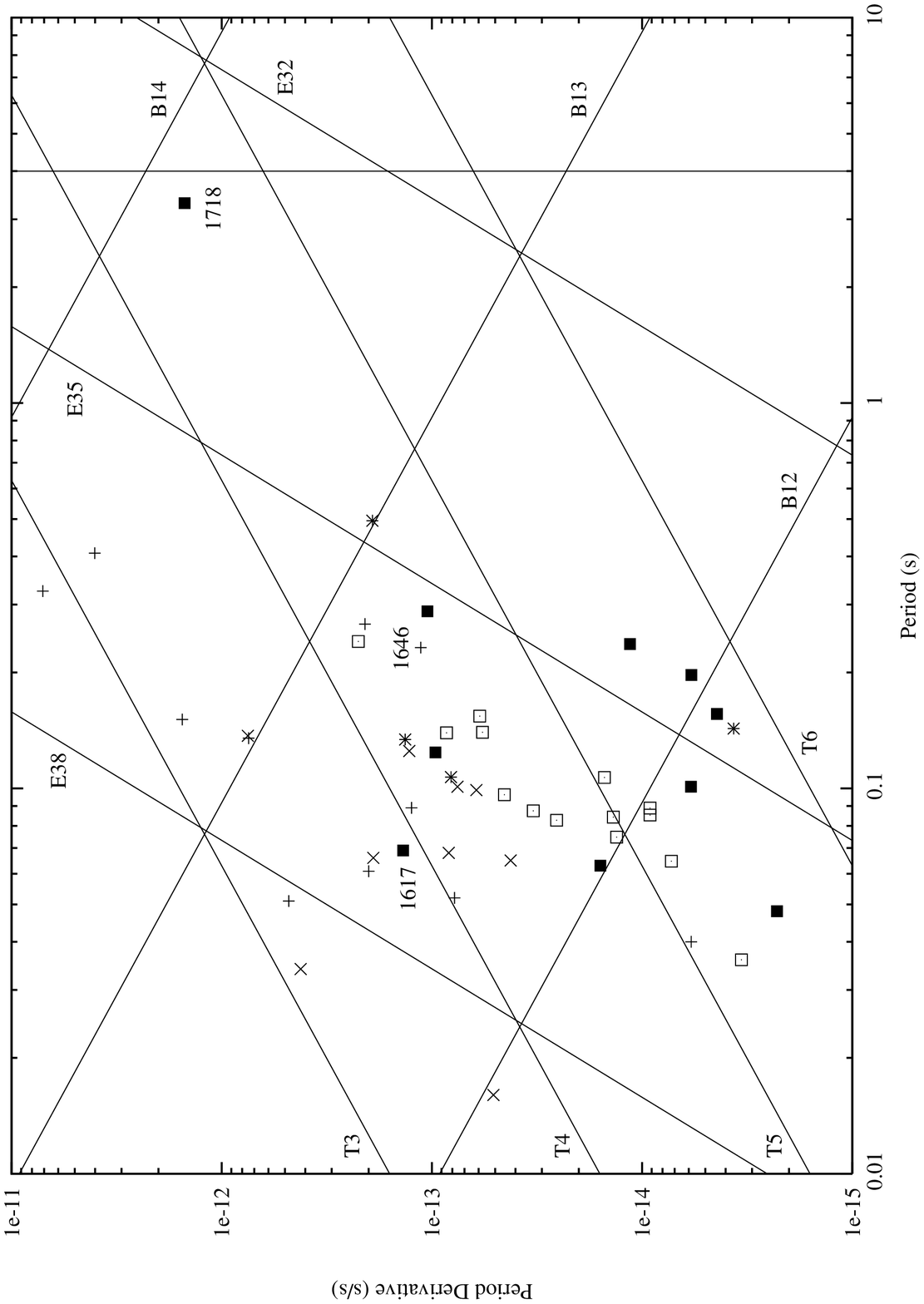}
\end{figure}

\end{document}